\documentclass[twocolumn,showpacs,preprintnumbers,amsmath,amssymb]{revtex4}

\usepackage{graphicx}
\usepackage{dcolumn}
\usepackage{bm}
\usepackage{subfigure}

\begin{document}


\title{ Study of proton capture resonant state of $^{15}$O at 7556 keV} 

\author{Sathi Sharma$^1$}
\author{Arkabrata Gupta$^2$}
\author{M. Roy Chowdhury$^3$}
\author{A. Mandal$^3$}
\author{A. Bisoi$^2$}
\author{V. Nanal$^3$}
\author{L. C. Tribedi$^3$}
\author{M. Saha Sarkar$^1$}
\email{maitrayee.sahasarkar@saha.ac.in}
\affiliation{$^1$Saha Institute of Nuclear Physics, HBNI, 1/AF Bidhannagar, 
Kolkata - 700 064, India}
\affiliation{$^2$Indian Institute of Engineering Science and Technology, 
Shibpur, Howrah - 711 103, India}
\affiliation{$^3$Department of Nuclear and Atomic Physics, Tata Institute of 
Fundamental Research, Colaba, Mumbai - 400 005, India}

\date{\today}

\begin{abstract}

 The slowest reaction in the  CNO cycle $^{14}$N(p,$\gamma$)$^{15}$O  has 
been studied by populating the E$^{lab}_{p}$=278 keV (E$^{r}_{c.m.}$=259 
keV) proton capture resonant state of $^{15}$O at 7556 keV. The strength 
of the resonance ($\omega\gamma$)  has been determined from the 
experimental data.
The level lifetime of the sub-threshold resonant state at E$_{x}$=6792 
keV, as well as the lifetimes of the 5181 keV and 6172 keV states, have 
been measured using the Doppler shift attenuation method (DSAM).   The 
structural properties of the nucleus $^{15}$O, such as, the level energies, transition 
strengths, level lifetimes, and spectroscopic factors, have been 
calculated theoretically by using the large basis shell model, which 
agrees reasonably well with the present as well as the previous experimental 
data.
\end{abstract}

\pacs{25.40.Lw, 26.20.+f}
\maketitle

\section{Introduction}
The study of nuclear reactions relevant to the nucleosynthesis process is very important to trace the origin and evolution of the different 
elements and to provide a stringent test to theoretical models \cite{iliadis}. 
In the first evolutionary stage of stars, the energy production mainly occurs via hydrogen burning inside the core through the PP (proton-proton) chain.
The low metalicity population \textrm{II} stars with the mass M$>$1.5 M$_{solar}$, initiate hydrogen burning via the CNO (carbon-nitrogen-oxygen)
 cycle along with the PP chain. 
The CNO cycle reactions have the same end-product, \textit{i.e.}, $^4$He as that for the PP chain. The   C,
N, O, or F nuclei act only as catalysts, as their total abundances are not altered while only the hydrogen is consumed. 
Due to the highest Coulomb barrier ($\sim$ 
2460 keV) of the $^{14}$N(p,$\gamma$)$^{15}$O reaction, it is the slowest reaction or the bottleneck of the CNO cycle \textrm{I}. So, 
this reaction plays a vital role in the stellar energy production and the reaction rate determination. 
The CNO cycles \textrm{II}, \textrm{III}, and \textrm{IV} do not contribute much to the energy production due to the higher Coulomb barrier. The globular clusters provide a strict 
constraint for the stellar models \cite{iliadis} because of their distinct
features in the Hertzsprung Russell [H-R] diagram.  
The age of the globular cluster can be determined from the main sequence 
(MS) turn off point, which is related to the $^{14}$N(p,$\gamma$)$^{15}$O reaction rate. The lower limit of the
age of the universe is also estimated from the globular cluster age 
determination \cite{chaboyer}.
There are many other implications of the $^{14}$N(p,$\gamma$)$^{15}$O 
reaction which are 
discussed elaborately in Refs. \cite{iliadis,chaboyer,galinski,daigle}.

Schr{\"o}der {\it et al.} \cite{schroder} studied the 
$^{14}$N(p,$\gamma$)$^{15}$O direct capture reaction in the energy range of E$_{p}$=200-3600 keV. After 
analyzing the cross-section 
data, they suggested a significant contribution of the E$_{c.m.}$=-504 
keV sub-threshold resonance state at 6792 keV of $^{15}$O in the total astrophysical S-factor at zero energy, \textit{i.e.}, S(0) value. They had kept the $\gamma$- ray width 
($\Gamma_{\gamma}$) as a free parameter in the theoretical fit and obtained the width as 6.3 eV. The R - matrix analysis was performed by Angulo {\it et al.} \cite{angulo} with all the previously measured data.
They got a completely different value of the radiative width of the 6792 
keV sub-threshold resonant state. The value of $\Gamma_{\gamma}$ was 3.6 
times less 
than the value mentioned by Schr{\"o}der {\it et al.} \cite{schroder}. 
The discrepancy between the value of radiative width ($\Gamma_{\gamma}$) 
motivated experimentalists and theoreticians to do new experiments and R - matrix fits.

If an excited state decays  with nearly 100\% probability via emission of $\gamma$- rays, then it is possible to obtain the total radiative width $\Gamma$ ($\simeq$ $\Gamma_{\gamma}$) by measuring the lifetime ($\tau$) of the
state. They are inversely related via the relation, 
$\Gamma$=$\hbar$/$\tau$. 
Bertone {\it et al.} \cite{Bertone} measured the lifetimes of the 5181 keV, 
6172 keV and 6792 keV states of the nucleus $^{15}$O using the Doppler shift attenuation method (DSAM). The measured lifetime of the 6792 keV state was 1.60$^{+0.75}_{-0.72}$ fs, which included the statistical uncertainties. 
However, if they had used the implanted target density corresponding to the compound TaN instead of the Ta, the change in stopping power would have resulted
in a higher value of lifetime, 3.2$\pm$1.5 fs. In Ref. 
\cite{schurmann}, Sch{\"u}rmann 
{\it et al.} had used the same centroid shift method with data at eleven different angles with better statistics. 
But they could only set an upper limit in the lifetime value ($\tau$ $<$ 0.77 fs). The latest value of the experimental lifetime of the 
6792 keV state was obtained by Galinski {\it et al.} \cite{Naomi} using the inverse kinematics. 
They have used the
$^{3}$He($^{16}$O,$\alpha$)$^{15}$O reaction to populate the excited 
state of 
$^{15}$O at a beam energy of 50 MeV with maximum 
recoil velocity $\beta$=0.05. As the maximum recoil velocity is 5$\%$ of the speed (c) of light in vacuum, a Monte Carlo code was written 
using the relativistic kinematics to obtain the lifetime value. Using the maximum likelihood method, they got an upper limit of
 $\tau$ $<$ 1.84 fs, which corresponds to a lower bound on the width, \textit{i.e.}, $\Gamma$ $>$ 0.44 eV. 
Therefore in Ref. \cite{Naomi}, they gave only the upper limit of the lifetime of the 6792 keV state using the data only at one angle, 
which was at 0$^{o}$.
Thus,  the lifetimes determined using the centroid shift method described 
in Refs. \cite{schurmann,Naomi} gave different results. So, uncertainty remains in the central value of the lifetime of the sub-threshold resonant state.

 In our present work, we have therefore measured the lifetime of the sub-threshold state by populating it via $^{14}$N(p,$\gamma$)$^{15}$O resonance reaction at lab energy 278 keV.
We have used the DSAM analysis (with the non-relativistic kinematic 
equations), as discussed in Refs. \cite{Bertone,schurmann}.
 The lifetimes of the 5181 keV, 6172 keV states of the nucleus $^{15}$O have also been measured in this experiment. The resonance strength ($\omega\gamma$) of the 278 keV narrow resonance is determined using the experimental data. 

The rate of the $^{14}$N(p,$\gamma$)$^{15}$O reaction is a sensitive function of the structural properties of the $^{15}$O nucleus.
These structural properties should be obtained from experiments and validated by theory for more accurate inputs to the nuclear astrophysics models. Many experiments were performed to measure the astrophysically relevant properties of the $^{15}$O nucleus; however, very few theoretical calculations were done so far.
Thus we have also done  large basis shell model calculations to study the  
low lying energy levels, level lifetimes, proton 
spectroscopic factors of $^{15}$O nucleus up to the resonance state at 
7556 keV.
The calculations have been performed using NuShellX code \cite{nushellx}.
The theoretical results for the excitation spectra, transition probabilities,  
level lifetimes and spectroscopic factors 
are compared with the present and other available experimental data 
\cite{schroder}. 

\section{EXPERIMENTAL DETAILS}

\subsection{The implanted target}
\label{target}

One of the most effective technique to produce targets which are
isotopically pure and can withstand high beam load over a long time is 
implantation technique \cite{seuthe,lee}. $^{14}$N$^{3+}$ ions of energy 75 keV from the ECR (Electron Cyclotron Resonance) ion source of the 
low energy ion accelerator \cite{lokesh1,lokesh2,lokesh3,lokesh4}  at Tata Institute of Fundamental Research (TIFR), Mumbai,  were implanted into 0.30(5) mm thick Ta backing. Tantalum (Ta) was chosen as the backing material due to its low sputtering rate, high saturation value than other 
materials like Au, Cu, $\textit{etc}$. The sputtering yield was simulated using TRIM simulation 
software \cite{trim}. The implantation has been done with a dose of  7.8$\times10^{17}$ atoms/cm$^{2}$. The beam was uniformly rastered over the Ta surface to have a uniform implantation circular zone with 2.5 cm diameter. Several techniques like X-ray photoelectron spectroscopy (XPS), scanning electron microscopy (SEM), secondary ion mass spectrometry (SIMS) have been used for surface characterization of the implanted target before and after implantation. The Ta backing was cleaned using ethanol and argon flashing before implantation to 
remove impurities in the backing. The SIMS analysis after 6 hours of sputtering with 5 keV Cs ion, indicated that C, F, and 
Na contamination sharply decreased with increasing depth. However, oxygen impurity existed deep inside the backing. All the techniques discussed so far are reported
in Refs. \cite{abhijit1,abhijit2,abhijit3,epj}. 

\begin{figure}
	\centering
	\includegraphics[width=9.5cm,height=7.5cm]{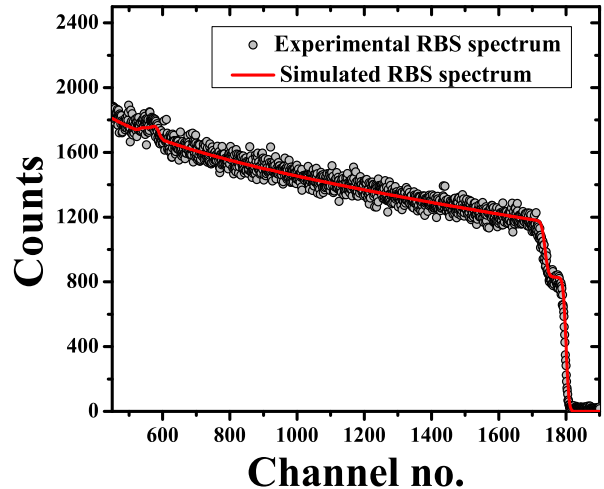}
	\caption{(Color online) A typical RBS spectrum of the implanted target. The SIMNRA fit 
\cite{simnara}
is shown in the solid red line.}
	\label{rbs}
\end{figure}

The bulk characterization of the implanted target is  done by Rutherford 
backscattering spectrometry (RBS) to get a quantitative estimation of the   stoichiometry of the implanted 
target \cite{abhijit3,epj}. $^{4}$He$^{2+}$ 
ion beam with 3.65 - 3.70 MeV energy and  12.2 nA current from 1.7 MV 
Pelletron accelerator at Inter-University Accelerator Centre, New Delhi, was used to do the measurement. The scattered He ions were detected by a Si surface barrier detector placed at 165$^{o}$ with respect to the beam direction. The RBS spectra were acquired for both  the $^{14}$N implanted  
Ta and the bare Ta. The experimental spectra for the scattered ions from 
the implanted target bombarded  with He ions at
3.682 MeV was fitted with the SIMNRA package \cite{simnara} with stopping 
power from  SRIM \cite{trim} as inputs. The best fit with reduced $\chi^{2}$ value 1.0, was obtained with N: Ta concentration ratio as 3:2. The concentration ratio was determined with an uncertainty of less than 1$\%$. 
The fitted plot is shown in Fig. \ref{rbs}. The Ta/N ratio for the 
implanted target is 0.667(33), which is similar to the target used by 
S. Daigle {\it et al.} \cite{daigle} in a previous work.
\subsection{Experimental setup and procedure}
The $^{14}$N(p,$\gamma$)$^{15}$O resonance experiment was performed at 
TIFR, Mumbai using ECRIA (ECR based ion accelerator) as mentioned in section \ref{target}.  The typical beam current on target during 
the experiment was 3 - 4 $\mu$A. The energy spread of the beam was 2 - 3 keV. The target was mounted at the end flange of the 0$^{o}$ 
beam line. As the beam current was not so high, there was no additional arrangement for cooling the target. The total charge collected during each 
run was measured using a current integrator. 

Two detectors were used in the experiment.  One of them was an electrically cooled CANBERRA (Mirion) Broad Energy Ge (BEGe) Falcon 5000 detector. The other one was a  p-type LN$_{2}$ cooled High Purity Germanium (HPGe) detector from Baltic Scientific Instrument. The BEGe detector was cylindrical with a 3 cm radius and 3 cm in length with 18\% relative efficiency. The  LN$_{2}$ cooled HPGe detector was also cylindrical with a 2.7 cm radius and 6.3 cm in length with 30\% relative efficiency.

Both the detectors were placed at an angle of 90$^{o}$ to calibrate the in-beam spectra of the detectors. Standard radioactive source $^{152}$Eu was used to calibrate the energy and efficiency of the detectors up to energy 1.5 MeV. For higher energy  up to 7 MeV,  calibrations were performed by using 
the  in-beam  
$\gamma$- rays emitted by the resonance state of  $^{15}$O \cite{falcon 
paper}.  

While scanning the implanted target, the BEGe detector was placed at 1.7 cm from the target center at $0^o$ to maximize the efficiency of the detector.  The proton beam energy varied from 278 keV to 312 keV in steps of 3 keV each.  The total charge accumulated in each run was estimated using a charge integrator.

\begin{figure}
	\centering
	\includegraphics[width=10.0cm,height=7.5cm]{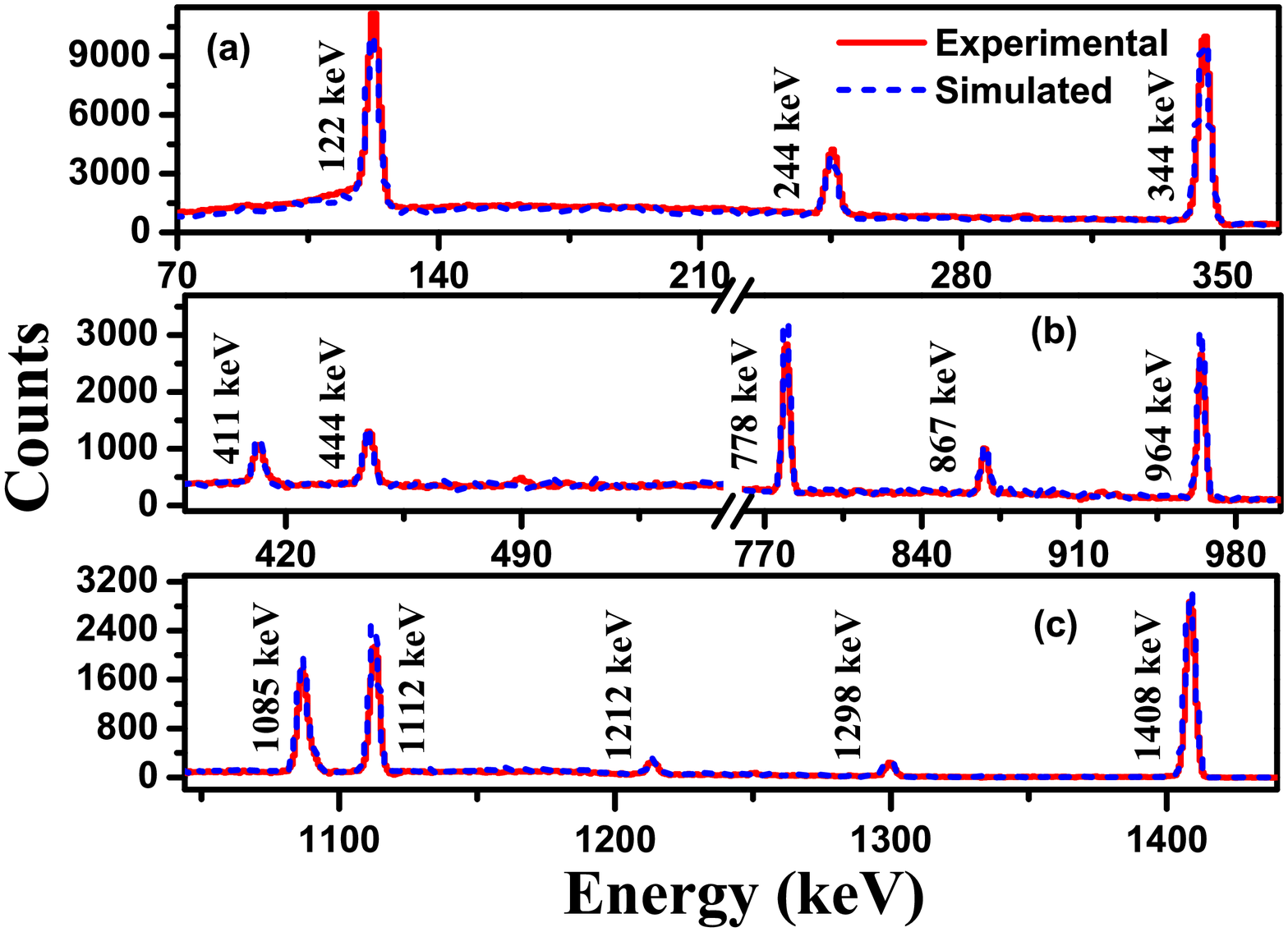}
	\caption{(Color online) Comparison of the experimental and simulated energy 
spectra of $^{152}$Eu for HPGe detector at 90$^{o}$.  (a) Low  (70-365 keV) (b) medium (390-1000 keV), and (c) high (1044-1440 keV) energy regions of the spectra are shown. The red solid line and 
blue dashed line connect the experimental and simulated data points, respectively.}
	\label{152eu}
\end{figure}

However, for acquiring DSAM data for lifetime measurement, the proton energy was kept fixed at 293 keV. The BEGe detector was kept at 90$^o$ as well as at 0$^{o}$, 25$^o$, 50$^o$, and 70$^o$ with respect to the beam direction. The other HPGe detector was placed at  
90$^{o}$ as well as 120$^{o}$ and 137$^{o}$ to acquire lifetime measurement data. 
Both the detectors were placed at 5 cm from the target center. As the end flange was 1.3 cm thick, so the distance between the target and the detector was 6.3 cm. Both the detector responses were measured experimentally and 
compared with 
GEANT4 simulation \cite{geant4}. The beam spot had a shift  from the 
central position in the end-flange  
(the position of the $^{152}$Eu source during off-beam measurements). We have used the simulation to match the $^{152}$Eu and in-beam resonance data for efficiency calibration of the detector over the whole energy range. The simulated data have been first matched with the experimental data, including the exact experimental conditions (the target position, 
the target holder geometry, $\textit{etc}$.). Later absolute efficiency of the detector is simulated for point sources. Thus,  the effects of the target holder asymmetries, as well as source position mismatch, were eliminated.
The relative and absolute efficiencies of the detectors were measured and compared with simulation. The GEANT4 simulation matched the experimental results within 1$\sigma$ level of uncertainty. The characterization of  the BEGe detector until 7 MeV has been  described 
in details in Ref. \cite{falcon paper}.
The response of the  HPGe detector was also obtained similarly. 
A comparison of the experimental and  simulated energy spectra for the 
HPGe detector at an angle 90$^{o}$ is shown in Fig. \ref{152eu}.\\

The experimental data of the BEGe detector was acquired using GENIE-2000 \cite{genie} data acquisition system (DAQ). 
It has a comprehensive set of capabilities for pulse processing, 
acquiring data and analyzing spectra from Multichannel
Analyzers (MCAs). MCA control, spectral display and manipulation, spectrum analysis, and reporting \cite{genie} are the basic functionalities of the DAQ system. The inherent gain of the setup was set to a minimum to acquire the $\gamma$- ray data till $\simeq$  8 MeV. 
The data were taken in 8k channels and singles mode. The analysis was done using GENIE-2000 \cite{genie} and an analysis software INGASORT \cite{Ranjan}. For the HPGe detector, we  had used a CAEN 
DT5780M (16 channel, 100 MS/s, 14 bit) digitizer \cite{caen} for pulse processing and data acquisition. The spectra were acquired in 16k channels and singles mode with minimum gain to have the data till 8 MeV to reduce the error in centroid determination. 

\section{Results and Analysis}
\begin{figure}
	\centering
	\includegraphics[width=9.8cm,height=8.0cm]{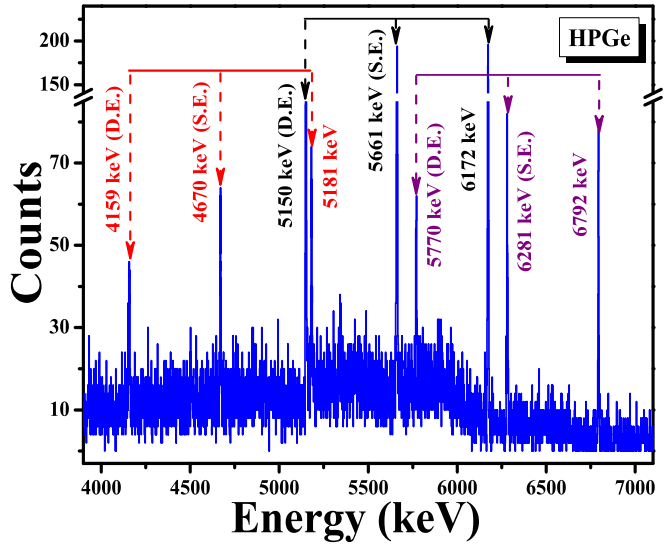}
	\caption{(Color online) Typical high-energy part of $\gamma$- ray energy spectrum 
from $^{14}$N(p,$\gamma$)$^{15}$O resonance
reaction at E$^{lab}_{p}$=278 keV of HPGe detector at an angle of 
90$^{o}$. S.E. and D.E. denote single escape and double escape peaks, 
respectively.}
	\label{hpge_spec}
\end{figure}

To calibrate the in-beam spectrum for both the detectors, we have used the data at 90$^{o}$ angle.
The high energy part of the spectrum of the HPGe detector at 90$^{o}$ is 
shown in Fig. \ref{hpge_spec}. 
The strong $\gamma$- rays 
coming from room background (like 511 keV, 1460 keV, and 2615 keV) and 
$\gamma$- rays emitted from the excited state (7556 keV) of $^{15}$O nucleus have been used for calibration. The gain stability of the detectors and electronics were assured by continuous monitoring of the background $\gamma$- lines.
Two crucial experimental information have been obtained from the present data. They are -- 
\begin{itemize}
\item{} the strength of the 259 keV resonance (7556 keV excited state) in $^{15}$O populated by proton capture resonant reaction and,
\item{} the lifetimes of the excited states of $^{15}$O populated in the 
present resonance reaction by analyzing the centroid shifts of the 
associated 
$\gamma$- rays due to Doppler shift. 
\end{itemize}

\subsection{Resonance Strength}
\subsubsection{Yield curve analysis}
\begin{figure}
	\centering
	\includegraphics[width=9.5cm, height=7.5cm]{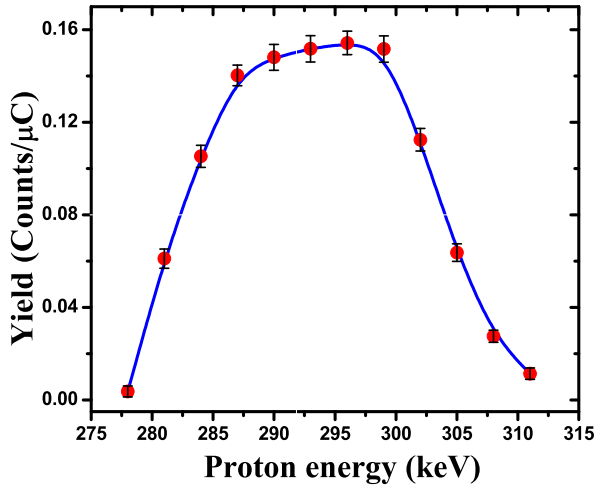}
	\caption{(Color online) Gamma ray yield plot of the E $^{lab}_{p}$=278 keV 
resonance of $^{14}$N(p,$\gamma$)$^{15}$ reaction.
The solid blue line is a guide to the eye.  }
	\label{yield}
\end{figure}

As already discussed in the last section, while scanning the implanted target, proton energy was varied from 278 keV to 312 keV in steps of 3 keV. The BEGe detector was placed at 0$^{o}$ at a distance of 1.7 cm from the target center. The 
strongest primary $\gamma$- ray of $^{15}$O 
compound nucleus in $^{14}$N(p,$\gamma$)$^{15}$O resonance reaction is 1384 keV. The yield of the reaction at a particular 
incident energy is  given by,
\begin{equation}
Y = \frac{N_{reaction}}{N_{beam}}=\frac{N_{peak}}{BN_{beam}W\eta_{peak}}
\end{equation}
where $N_{reaction}$ is  the total number of reactions that occurred and 
$N_{beam}$ is  the total number of
incident projectiles.  B, $N_{peak}$, W and $\eta_{peak}$ are the 
branching ratio (probability of emission of that particular $\gamma$- ray
per reaction), the total number of photons emitted by the state excited 
by the reaction (given by the area under the corresponding photo-peak),  
the angular correlation, and the detector efficiency, respectively, for a specific nuclear transition. The yield has been determined by utilizing the area under the 1384 keV $\gamma$- ray photo-peak in the spectrum at each beam energy, incorporating other necessary factors. The yields have been plotted 
as a function of incident proton beam energies (Fig. \ref{yield}) to generate the yield curve.  In the present yield plot, we got a plateau region from 288 keV to 300 keV.  To compare the experimental yield profile with TRIM simulation \cite{trim}, we have expressed the incident proton beam energy in terms of the linear thickness of the target,  using SRIM stopping power \cite{trim}. The experimental profile matches with simulation considering the density of the target similar to Ta only. The analysis of the target scanning results has been discussed in detail in Ref. \cite{epj}. From the yield plot, the measured energy thickness ($\Delta$E) of the implanted target was $\simeq$ 21$\pm$1 keV.

For a thick target,  whose energy width is more than the energy width of the resonance,  the resonance strength can be determined using the height of the plateau region in the yield curve.  The resonance strength 
($\omega\gamma$) and maximum yield for a thick target is related through 
the equation,
\begin{equation}
\label{strength}
\centering
\omega\gamma = \frac{2\epsilon_r}{\lambda^{2}_{r}} Y_{max, \Delta 
E\rightarrow\infty}, 
\end{equation}
where, $\lambda^{2}_{r}$ is the corresponding de Broglie wavelength and $\epsilon_r$ is the effective stopping power at the resonance energy. 
 For determining the resonance strength, effective stopping power at the resonance energy ($\epsilon_r$) for Ta has been used in the present work.

As the energy thickness (${\Delta E}$) of our target is nearly 21 keV and 
the width ${\Gamma}$ of the resonance is 0.99(10) keV \cite{nndc}, the 
ratio ${\Delta E}/{\Gamma}$ is $\approxeq$ 21. It has been shown in 
\cite{iliadis} that if the target thickness is $\simeq$ 20 times larger 
than 
the total resonance width ($\Delta E/\Gamma \simeq$ 20), the maximum 
yield at the plateau is $\simeq $ 95\% of the
yield for an infinitely thick target and the full width at half maximum 
(FWHM) of the yield curve is equal to the target thickness within 0.5\%. The uncertainties in the resonance strength are both statistical as well as systematic. The sources of systematic errors are the amount of  total 
the charge accumulated, target stoichiometry, the effective energy, and branching of the corresponding 
$\gamma$- ray $\textit{etc}$. Including the uncertainties, the present value of the resonance strength is, 
$\omega\gamma$=12.78$\pm$0.29(stat.)$\pm$0.92(sys.) meV. The present experimental value has been compared with the previous  values from literature in Table \ref{strengths}. The weighted average value of the resonance strength ($\omega\gamma_{average}$) is  12.7 (2) meV.
\\

\begin{table}
\caption{\label{strengths} Comparison of experimental resonance strength ($\omega\gamma$ (meV))
determined in the present work with earlier literature values. The  
uncertainties are quoted within brackets. The weighted average value is also shown.}
\begin{ruledtabular}
\begin{tabular}{ccc}
Reference& \multispan{2} \hfil $\omega\gamma$ (meV)\hfil\\
&Measured &Average\\
 \hline\hline
Present work &  12.8 (9)\\
Ref. \cite{daigle} & 12.6(3)\\
Ref. \cite{bemmerer} & 12.8(6)\\
Ref. \cite{Imbriani} & 12.9(9)&12.7(2)\\
Ref. \cite{Runkle} & 12.4(9)\\
Ref. \cite{Becker} & 13.7 (10)\\
\end{tabular}
\end{ruledtabular}
\end{table}

\subsection{Lifetime measurement}
\begin{figure}
	\centering
	\includegraphics[width=9.5cm, height=7.5cm]{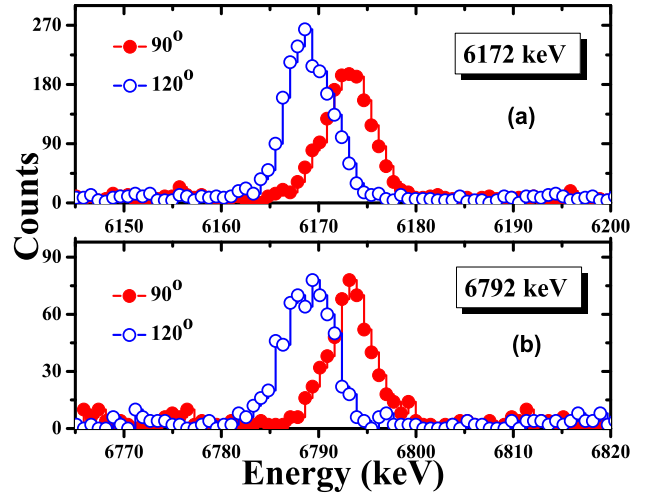}
	\caption{(Color online) Full energy peaks for (a) 6172
keV and (b) 6792 keV $\gamma$- rays at two different angles of the HPGe detector. The red filled circles joined by red line and
blue open circles joined by blue line  correspond to the spectra at 90$^{o}$ and 120$^{o}$, respectively.}
	\label{spectrum}
\end{figure}

When $\gamma$- rays of particular energy are emitted from a recoiling nucleus, while it slows down through the target medium, their energies are shifted. The shifted energy depends on the initial recoil velocity 
$(v_{0})$, angle between the detector detecting the $\gamma$- ray  and the recoiling nucleus $(\theta)$, velocity attenuation factor $(F(\tau))$  and the correction factor $(P)$ for the finite size of the detector. The attenuation factor, $(F(\tau))$, is a function of the lifetime of the nuclear level emitting the $\gamma$- ray and the recoiling  medium.
The relation between the energy of the  $\gamma$- ray detected by the 
detector at an angle, $\theta$, designated as $(E_\gamma^\theta)$ and 
actual energy of the $\gamma$- ray  $(E_\gamma^o)$ is given by,
\begin{equation}
\label{ftau_expt}
\centering
E_\gamma^\theta=E_\gamma^o \left[1 + \beta_0 F(\tau) Pcos\theta\right]
\end{equation}
where, $\beta_0= {v_{0}}/{c}$.  
The spectra acquired by the HPGe detector at 90$^{o}$ and 120$^{o}$ for 
6172 keV and 6792 keV $\gamma$- rays
are shown in Fig. \ref{spectrum}.

\begin{figure}
	\centering
	\includegraphics[width=9.5cm,height=7.5cm]{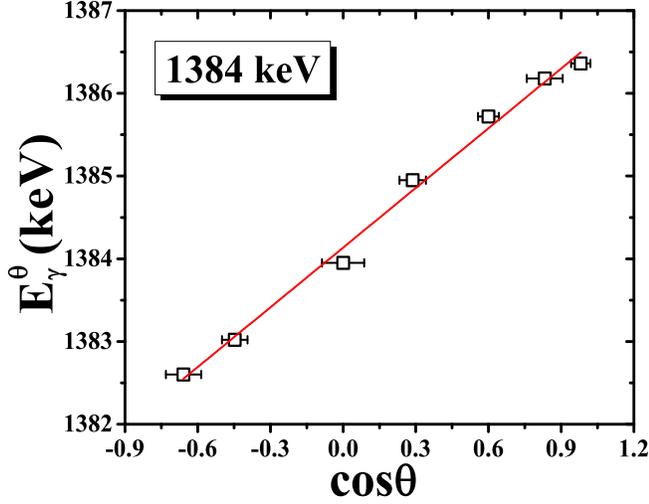}
	\caption{(Color online) Doppler shifted $\gamma$- ray energy 
(E$_{\gamma}^{\theta}$) of 1384 keV $\gamma$- line plotted against 
cos$\theta$, where $\theta$ is the detection angle. The solid red line corresponds to fit according to 
Eq. \ref{ftau_expt}.
}
	\label{betap_1stjune}
\end{figure}

\begin{figure}
	\centering
	\includegraphics[width=9.7cm, height=7.5cm]{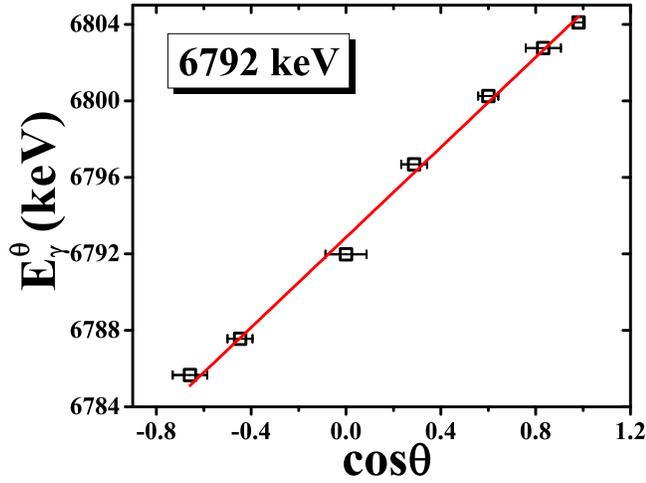}
\caption{(Color online) Doppler shifted $\gamma$- ray energy (E$_{\gamma}^{\theta}$) of 
6792 keV $\gamma$- line plotted against cos$\theta$, where $\theta$ is the detection angle. The solid red line corresponds to fit according to 
Eq. \ref{ftau_expt}. 
}
	\label{6792ftau}
\end{figure}

 The reported energy width of the 7556 keV state (259 keV resonance 
state)  is 0.99(10) keV \cite{nndc}. The associated lifetime of the state is thus 0.66$\times 10^{-18}$ s. This lifetime is too small compared to the stopping time of the recoiling nucleus in the medium, and thus the recoil velocity is not 
attenuated, \textit{i.e.}, F$ (\tau)$=1. For this reason, the lifetime of the resonance state could not be determined by the DSAM technique. However, the shifts of the primary $\gamma$- rays emitted from this level have been utilized to 
determine the factor of $\beta_0P$. The shifts of 1384 keV $\gamma$- ray at different angles are plotted with cos$\theta$, as shown in Fig. \ref{betap_1stjune}. The 
factor $\beta_0P = 0.001738 \pm 0.000343 $ has been obtained from a linear fit to the data points considering F$(\tau$)=1.
Next, the DSAM method is used to determine the lifetimes of the lower excited states of $^{15}$O, from where
 the secondary $\gamma$- rays originate. The centroids of three secondary $\gamma$- rays 5181 keV, 6172 keV, and 6792 keV are 
determined at seven different angles. The shifted centroid values are plotted against cos$\theta$. The angle ($\theta$) values have been
 corrected to account for the shift of the beam spot from the center of the target flange. The plot for the  6792 keV $\gamma$- ray 
is shown in Fig. \ref{6792ftau}. The data points have been fitted using Eq. \ref{ftau_expt}. The F($\tau$) value for individual $\gamma$- ray
energy (see Table \ref{tau}) has been determined after including the factor $\beta_0P$ which was obtained from the linear fit shown in Fig. \ref{betap_1stjune}.

In general, the direction of motion of the recoil given by $\beta(t)$ at a particular time t will differ from that of  $\beta(0)$ (at t=0) due to the scattering of the recoil nuclei as they are losing energy in the target medium. These changes in the direction should be included in the definition of $F(\tau)$. The  instantaneous angle of $\beta(t)$ to the
beam direction axis (z-axis, say) is $\phi(t)$, such that 
$\beta_z(t)=\beta(t) cos\phi(t)$, and $\beta_z(0)\equiv\beta(0)$. 
$F(\tau)$ is an attenuation coefficient which lies between 0 and 1.
The lifetime of the $\gamma$- emitting leve1 can be determined if $F(\tau)$ differs from 0 and 1; otherwise, a limit can be obtained for the mean lifetime. The velocity attenuation coefficient F($\tau$) is expressed as,

\begin{equation}
\label{ftau_theo}
\centering
F(\tau)=\frac{1}{v_{0}\tau} \int_{0}^{\infty}\overline{v(t)cos\phi} ~e^{-
{t}/{\tau}}dt
\end{equation}

\begin{figure}
	\centering
	\includegraphics[width=9.8cm, height=7.5cm]{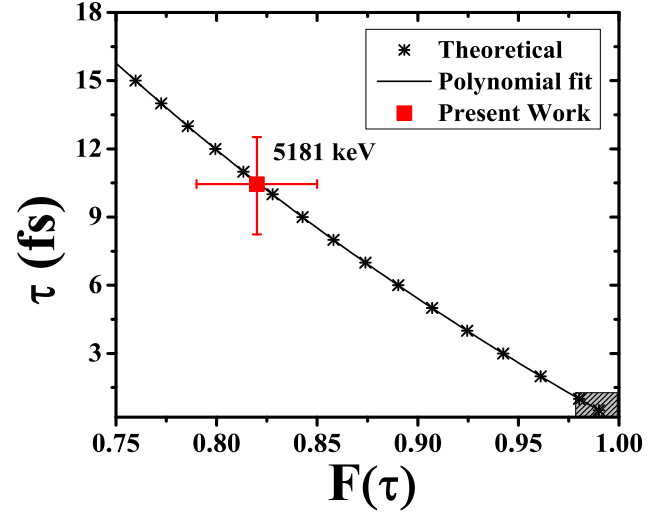}
	\caption{(Color online) The solid black line corresponds to F($\tau$) vs. $\tau$ 
curve according to Eq. \ref{ftau_theo}. 
The theoretical data points have been obtained from Eq. \ref{ftau_theo} with 
the help of TRIM software \cite{trim}.
The grey color shaded area corresponds to the allowed region for the 6172 
keV and 6792 keV $\gamma$- rays for the present work.
The red square corresponds to the 5181 keV $\gamma$- ray. The errors 
associated with the F($\tau$) and $\tau$ values for the 5181 keV 
$\gamma$- ray has also been shown in the figure.} 
	\label{ftau_1stjune}
\end{figure}

\begin{figure}
	\centering
	\includegraphics[width=10.0cm, height=8.0cm]{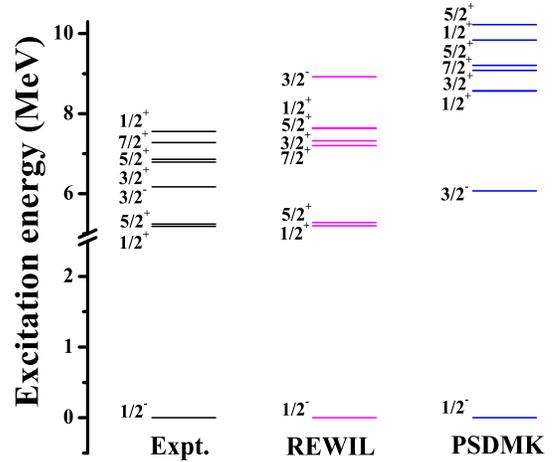}
	\caption{(Color online) Comparison of experimental energy levels with theoretical 
calculations of $^{15}$O.}
	\label{level}
\end{figure}

Here, $\tau$ is the mean lifetime of the excited states, $v(t)$ is the velocity of the recoil nuclei at time t, $\phi$ is the scattering angle, and  $\overline{v(t)cos\phi}$ is the time-dependent averaged projection of the recoil velocity distribution. The $z$ component of ion velocity is changing with time as functions of the characteristic slowing-down time of the ions due to electronic processes, the initial velocity of the ions $v(0)$, the electronic and the nuclear stopping and scattering parameters. 
In the present work, stopping powers are taken from the SRIM 2013 
\cite{trim} software. 
Although stoichiometry of the implanted target is not of pure Ta, the 
observations discussed
in Ref. \cite{epj} indicate that  N occupies interstitial, rather than 
substitutional
sites within the Ta lattice as also observed by earlier workers 
\cite{Bertone, schurmann}. Therefore,  the density of the target is taken to be the same as that of pure Ta for the calculation. The value of $\overline{v(t)cos\phi}$ has been obtained using the collision details of a large number of the $^{15}$O recoiling nuclei in the Ta backing. 
The collision details of $^{15}$O ions have been calculated using the TRIM software \cite{trim}.
Then, the recoil velocity distribution was fitted with a sixth-order polynomial function. Next, Eq. \ref{ftau_theo} has been solved by replacing the 
$\overline{v(t)cos\phi}$ with the fitted polynomial function.
The theoretical F($\tau$) values for various values of $\tau$ have been 
calculated. As we have calculated experimental $F(\tau)$ values,
for convenience of extracting corresponding $\tau$ values, $F(\tau)$ 
values are plotted as an independent variable in Fig. 
\ref{ftau_1stjune}. 
The  curve is fitted by a 
fourth-order polynomial with 95$\%$ confidence limit which expresses 
$\tau$ as a function of $F(\tau$). 
The mean lifetime value of the 5181 keV state has been
obtained from the F($\tau$) vs. $\tau$ fitted curve
with the corresponding experimental F($\tau$) value. In the case of 6172 keV 
and 6792 keV $\gamma$- rays, the experimental $F(\tau)$ values
are very close to 1. So, we can set  upper limits of the 
lifetime values for 6172 keV and 6792 keV levels.
 The experimental F($\tau$) and lifetime values are mentioned in Table 
\ref{tau}. 
The errors in the lifetime values originate from the uncertainties in the 
stopping power taken from SRIM \cite{trim}, 
angle measurement, target stoichiometry, $\textit{etc}$.     

\begin{table*}
\caption{\label{tau}Experimental  F$(\tau)$ and lifetime values obtained 
in the present work and their comparison with  previous results.}
\begin{ruledtabular}
\begin{tabular}{ccccccc}
 & & & &\multicolumn{1}{c}{Lifetime ($\tau$) in  fs}\\
  E$_{x}$& &\multispan{5}\hrulefill\\
(keV)&F($\tau$)&Present& Ref. \cite{Bertone}&Ref. \cite{schurmann}&Ref. 
\cite{azenberg}& Ref. \cite{Naomi}\\ 
\hline\hline\\
5181&0.82$\pm$0.03&10.45$^{+2.07}_{-2.21}$&9.67$^{+1.34}_{-1.24}$&8.40$\pm$1.00&8.20$\pm$1.00&--\\
6172&1.00$\pm$0.03&$<$1.22&2.10$^{+1.33}_{-1.32}$&$<$0.77&$\leq$2.5&$<$2.5\\
6792&0.99$\pm$0.02&$<$1.18&1.60$^{+0.75}_{-0.72}$&$<$0.77&$\leq$28&$<$1.8\\
 \end{tabular}
\end{ruledtabular}
\end{table*}
\begin{table}
\caption{\label{lifetime}Comparison of experimental level energies and 
lifetimes of $^{15}$O  with the shell model
predictions.}
\begin{ruledtabular}
\begin{tabular}{ccccc}
\multispan{3}\hfil Expt.\hfil&\multicolumn{2}{c}{Theo.}\\
\multispan{3}\hrulefill&\multispan{2}\hrulefill\\
{Energy }& \multicolumn{2}{c}{Lifetime $(\tau)$}&Energy&$\tau$\\
(keV)           &Present &Prev.\cite{nndc}& (keV)&  \\ 
\hline\hline\\
5181&10.45$^{+2.07}_{-2.21}$ fs& 5.7(7) fs&5192&0.82 fs\\
5240& --& 2.25 (21) ps& 5276& 1.2 ps\\
6172&$<$1.22 fs& $<$1.74 fs&8920& 1.5 fs \\
6792&$<$1.18 fs& $<$20 fs&7318& 0.07 fs\\
6859&--& 11.1(17) ps&7631& 30.46 ps \\
7276&--& 0.49(11) ps&7199& 0.42 ps\\
\end{tabular}
\end{ruledtabular}
\end{table}

\begin{table}
\caption{\label{tab:table1}Comparison of proton spectroscopic factors of 
previous experimental data with the shell model
predictions.}
\label{spectroscopic_factor}
\begin{ruledtabular}
\begin{tabular}{cccccc}
& & && \multicolumn{1}{c}{C$^{2}$S}\\
&&&Theory&\multispan{2}\hfil Expt \hfil\\
E$_{x}$&&&&\multispan{2}\hrulefill\\
(keV)& J$^{\pi}$&nl$_{j}$&Present& Prev.\cite{schroder} & 
Prev.\cite{Bertone2}\\
\hline\hline\\
0 & 1/2$^{-}$ & 1p$_{1/2}$       & 1.23& 1.29 (18)& 1.7(4)\\
5181 & 1/2$^{+}_{1}$ & 2s$_{1/2}$& 0.01& 0.004 (1)& 0.0049(15)\\
5240 & 5/2$^{+}_{1}$ & 1d$_{5/2}$& 0.1& 0.06 (1)& 0.094(20)\\
6172& 3/2$^{-}_{1}$ & 1p$_{1/2}$& 0.001& 0.038 (16)& 0.050(11)\\
6792& 3/2$^{+}_{1}$ & 2s$_{1/2}$& 0.96& 0.49 (1)& 0.51(11)\\
& & 1d$_{5/2}$& 0.004& - & 0.16(3)\\
6859 & 5/2$^{+}_{2}$ & 1d$_{5/2}$& 0.74& 0.37 (1)& 0.61(13)\\
7276& 7/2$^{+}_{1}$ & 1d$_{5/2}$& 0.99& 0.35 (1)& 0.66(14)\\
7556& 1/2$^{+}_{2}$ & 2s$_{1/2}$& 0.56& $\approx$ 0.49& 0.82(18)\\  
\end{tabular}
\end{ruledtabular}
\end{table}

\section{THEORETICAL CALCULATIONS}
\subsection{Partial wave analysis of the resonance}
One of the vital parameters of resonance reaction is the width of the unbound state. In the present work, the width of the 7556 keV resonant 
state has been determined theoretically. 
The single-proton width for the capture of a proton on a  quantum orbital defined by (nlj) is obtained from proton
scattering cross-sections calculated with a Woods - Saxon potential. The code WSPOT \cite{wspot} has been utilized to calculate the phase shift  and the 
scattering cross-section  as a function of (positive) energy. It searches for a resonance peak in an energy interval, and iteratively generates the full resonance curve. The  resonance energy peak and 
the full width at half maximum of the resonance (width of the 
unbound resonance state) are provided as outputs.  The parameters 
of the potential are chosen to have the best fit of nuclear
single-particle energies and nuclear radii. The program utilizes Woods -
Saxon potential as the phenomenological one body potential.
The parameters of the potential are chosen to have the best fit of nuclear
single-particle energies and nuclear radii \cite{rajan}. The different 
parameters used in the present calculation are 
V$_{o}$(central part)=-53 MeV, V$_{1}$(isospin dependent part)=-30 MeV 
and V$_{so}$=22 MeV for the potential
strengths. The radii parameters are  r$_{o}$(radius parameter-
central)=r$_{so}$(radius parameter-spin orbit)=1.25 fm and 
a$_{o}$(diffuseness-central)=a$_{so}$(diffuseness-spin orbit)=0.65 fm for 
diffuseness. The  radius for the Coulomb term is smaller with  
r$_{c}$=1.20 fm.
The variation of the energy of the incoming particle changes  the 
relative phase of the inner and outer
wavefunctions. The energy at which the amplitude of inside and outside
wavefunctions match, the cross-section has the maximum value. This energy is 
known as resonance energy.
Theoretically, the one proton 1/2$_{2}^{+}$ resonant state at 7556 keV 
is populated through a pure l=0, \textit{i.e.},  s wave capture. By varying the 
energy of the incoming proton, the width of the resonance is obtained.
 From the theoretical calculation, we get the width of the unbound state 
of 1.2 keV
which is close to the experimental width of 0.99(10) keV \cite{nndc}. 
The ratio of experiment over theory provides a measure of the spectroscopic factor, S = 0.82(9), which agree well 
with the reported experimental value of 0.82(18) \cite{Bertone2}. However, it deviates by a factor of   $\approx $ 0.6 from the value reported by Ref.  \cite{schroder} . 

\subsection{Large basis shell model calculations}

The rate of $^{14}$N(p,$\gamma$)$^{15}$O resonance capture reaction 
depends on the
structural properties of the low lying states in $^{15}$O.  
 As discussed in Ref. \cite{fortune}, the proton spectroscopic factors
 for states in $^{15}$O populated by  $^{14}$N($^{3}$He,d)$^{15}$O 
reaction and neutron spectroscopic factors for $^{15}$N populated by
 $^{14}$N(d,p)$^{15}$N reaction deviate by factors of 0.65 and 0.63 for 
l=0 and l=2 components, respectively, despite being mirror partners of 
each other. Fortune  \cite{fortune} addressed the problem and found that 
the actual l = 0 spectroscopic factors for two
3/2$^+$ states are significantly smaller than those recently reported. 
Thus, it is necessary to do 
theoretical calculations within the large basis shell model to extract the  
absolute spectroscopic strengths of the states of $^{15}$O.   

In the present work, we have used the code NuShellX \cite{nushellx} to do 
 large basis shell model (LBSM) calculations. 
For wavefunction and energy spectra calculation, ZBM model space 
\cite{nushellx} has been used. 
ZBM model space consists of 
$^{12}$C core and 1p$_{1/2}$, 1d$_{5/2}$ and 2s$_{1/2}$ as the valence 
orbitals. REWIL isospin interaction \cite{rewil} has been used 
for the calculations.
The energy spectra have been calculated till 20 MeV using full valence 
space, without any subshell restrictions.
The comparison of energy levels up to 8 MeV with the experimental data  
is shown in Fig. \ref{level}. 
All the positive and negative parity states are reproduced reasonably 
well except the 3/2$^{-}_{1}$ state, \textit{i.e.}, 6172 keV
state which is overpredicted (see Table \ref{lifetime}). 
The reduced transition probabilities for E2 and E1 transitions
have been calculated with effective charges e$_{p}$=1.35e and 
e$_{n}$=0.35e, respectively.
In the case of M1 and M2 transitions, standard values of intrinsic magnetic 
moments have been used.

The level lifetimes of the $^{15}$O nucleus have been calculated by using theoretical reduced 
transition probabilities and experimental $\gamma$- ray energy values and 
branching ratios, wherever 
needed. The lifetime values are compared with the experimental data from 
the present and previous work \cite{nndc}. 
Theory predicts the lifetime values quite well in most of the cases. \\

Another critical parameter for astrophysical model calculations is the
spectroscopic factor. The proton spectroscopic factors of the ground 
state, as well as the low lying states in $^{15}$O, have been determined. 
The squares of the overlap integrals, \textit{i.e.},  the spectroscopic factors
 have been calculated for $^{14}$N ground state with all the states of 
$^{15}$O up to the resonance state. The theoretical values have been compared with the experimental 
data in Table \ref{spectroscopic_factor}. In most of the cases, the calculated values are closer to the experimental data report in Ref. \cite{schroder}.
 However, for 6792 keV state, the calculated value  disagrees with the experimental data.  The absence of 1d$_{3/2}$ orbit in the model space may be one of the reasons for this discrepancy.
Interestingly the spectroscopic factor for 7556 keV from shell model agrees with the value reported in Ref. \cite{schroder} which deviate by a factor of $\approx$ 0.6 from
that reported by Ref. \cite{Bertone2} as well that predicted by partial wave analysis (Sec. IVA).

It has been discussed earlier that the energy of the 3/2$_{1}^{-}$ state is overpredicted and the 
spectroscopic factor is underpredicted.  To reproduce a negative parity state 3/2$_{1}^{-}$, only a single negative parity orbital 1p$_{1/2}$ 
 is present in the ZBM model space. 
The 1p$_{3/2}$ orbital is absent in the ZBM space, which may cause the 
discrepancy for 3/2$^-$ state.

Thus, another model space PSD has been considered with PSDMK interaction. 
The PSD model space, consists of 1p$_{3/2}$, 1d$_{3/2}$ as well as 
1p$_{1/2}$, 1d$_{5/2}$ and 2s$_{1/2}$ orbitals with a $^4$He core. In 
this case, full space calculations were beyond the present computational
capacity. Thus a suitable  truncation scheme has been adopted to perform 
the calculations. Sub-shell restrictions -- 
with six particles and two holes  in the 1p$_{3/2}$ orbital,
zero occupancies in the 1d$_{3/2}$ orbital and no restrictions to the other 
orbitals have been adopted. 
The ground state spin is reproduced, but the energy values are 
overpredicted (see Fig. \ref{level}).
The  ZBME model space (1p$_{1/2}$, 1d$_{3/2}$, 1d$_{5/2}$, 2s$_{1/2}$) 
with REWILE interaction have also been used. 
However, the changes in the energy eigenvalues are less than 1$\%$ 
compared to ZBM+REWIL calculations. The inclusion of 1d$_{3/2}$ 
orbital does not improve the results.\\

\section{SUMMARY AND CONCLUSIONS}

A few astrophysically important states of $^{15}$O were populated using  
$^{14}$N(p,$\gamma$)$^{15}$O resonance reaction at lab energy 278 keV 
using an
implanted target. The implanted target was characterized using standard 
techniques and its  stoichiometry was 
obtained using RBS data. 

To get depth profile of the implanted ions in the target, it was scanned 
with varying proton
energy from 278 keV to 312 keV. The strength of the resonance was  
evaluated using
thick target approximation as $\Delta$E $>>$ $\Gamma$ (using Eq. 
\ref{strength}). The effective
stopping power was calculated using SRIM 2013 \cite{trim} software. The 
Ta density is used to calculate the effective stopping power.
The calculated resonance strength is 12.8 (9) meV. It has been compared 
with previous measurements in Table \ref{strengths}. 
The weighted average value of the resonance strength ($\omega\gamma_{average}$)  has been found to be 12.7 (2) meV.

The lifetimes of the 6792, 6172 and 5181 keV states were measured using 
DSAM technique.
The centroid shift method was adopted to obtain the lifetimes
of the states. To calculate the velocity attenuation profile, we have 
used the target density of pure Ta.
In the present work, we were not able to determine finite lifetime values for 
the 6172 and 6792 keV states.  
The lifetime of the sub-threshold resonance state, \textit{i.e.}, 6792 keV was 
found to be $\tau$ $<$ 1.18 fs. 
So, the lower limit on the level width, $\Gamma$ is $>$ 0.56 eV. The measurement by 
Galinski {\it et al.} \cite{Naomi} gave the upper limit of the 
lifetime for 6792 keV state
 as $\tau$ $<$ 1.84 fs. So, the present measurement further constrained the 
lifetime value of the sub-threshold 6792 keV state.  
The obtained upper limit of the lifetime of the 6172 keV state is $\tau$ $<$ 1.22 fs. In 
case of 5181 keV state, we got a finite lifetime value
of 10.45$_{-2.21}^{+2.07}$ fs, which is in good agreement with the 
previous measurements \cite{Bertone,schurmann}. 

The partial wave analysis was used to calculate the resonance width of 
the  7556 keV state. The calculated  width of the 7556 keV state and its spectroscopic factor
are in good agreement with the literature values (Ref.\cite{Bertone2}).
The theoretical calculations using LBSM with ZBM model space and REWIL 
interaction reproduced the experimental data well in most of the 
cases. The resonance state at 7556 keV is reproduced theoretically at 
7646 keV using the shell model calculation. The calculated spectroscopic factor for 7556 keV state  agrees with that reported in 
Ref. \cite{schroder}. However, it disagrees with the data from Ref. \cite{Bertone2} and the calculated value from partial wave analysis. 
The lifetimes and spectroscopic factors for other observed states
are also calculated and compared with present and previous data, wherever 
available.
 However, some disagreements of  the theoretical results with 
experimental data  for a few states, indicate the need of improved 
interactions in the lighter mass region. 

\section{acknowledgements}

We would like to thank all the members of ECRIA lab at TIFR, Mumbai for 
their help and cooperation.

\end{document}